# What is it Like to Be a Bot:
Simulated, Situated, Structurally Coherent Qualia (S3Q) Theory of Consciousness

Schmidt, K., Culbertson, J., Cox, C., Clouse, H.S., Larue, O., Molineaux, M., Rogers, S.

> "…an organism has conscious mental states if and only if there is something that it is like to *be* that organism…" - Thomas Nagel

## ABSTRACT

A novel representationalist theory of consciousness is presented that is grounded in neuroscience and provides a path to artificially conscious computing. Central to the theory are representational affordances of the conscious experience based on the generation of qualia, the fundamental unit of the conscious representation. The current approach is focused on understanding the balance of simulation, situatedness, and structural coherence of artificial conscious representations through converging evidence from neuroscientific and modeling experiments. Representations instantiating a suitable balance of situated and structurally coherent simulation-based qualia are hypothesized to afford the agent the flexibilities required to succeed in rapidly changing environments.

## INTRODUCTION
Any theory of consciousness must explain the neural correlates, the behavioral utility, and the phenomenal characteristics of conscious experiences. Together these address the fundamental question: 'what is it like to be conscious and why is it like that?' The theory presented here suggests consciousness is a **simulated**, sensorimotor world model that is appropriately **situated** and **structurally coherent** to achieve acceptable stability, consistency, and usefulness. This theory is framed through neuroscience to propose experimental mechanisms that index the balance of representational flexibility and practicality in conscious systems.

S3Q Theory suggests conscious representations must suitably balance three tenets:

1. **Simulated**—The representation is an internally generated sensorimotor world model
2. **Situated**—All aspects of the representation are defined by relationships and interactions
3. **Structurally coherent**—The representation captures enough information about reality to facilitate interactions with the environment

All three of these tenets of consciousness can be investigated through experimentation and modeling of neural activity in the human brain. These activities involve communications between network components: **simulation** can be enabled by mechanisms such as synchronized neural feedback of internally generated models, **situatedness** can be enabled by mechanisms such as nested neural oscillations and lateral inhibition between processes, and **structural coherence** can be enabled by interaction of internally generated models and externally connected components for predictive learning. It is our contention that artificial intelligence agents must also balance these tenets of consciousness to achieve a flexible representation. Neuroscience has led to key advances in the development of novel AI technologies, and conscious processing in the brain can provide further insights into flexible AI through the translation of identified mechanistic principles of phenomenal experience (Rogers and Kabrisky, 1991).

**NEUROSCIENCE**
In the human brain, sensory inputs form into features represented by the central tendency of cell populations tuned to different values of that feature (Hubel and Wiesel, 1962). These inputs feed forward through a hierarchical representation which is then incorporated into feedback to earlier processing areas (Bullier, 2001). Synchronous neural activation patterns emerge from these hierarchical recurrent interactions between brain regions, which are characterized by distinct frequency bands of electrophysiological activity (Buzsaki, 2006). It is hypothesized that these synchronizing networks in the human brain offer key insights to the production of qualia, which are the fundamental computational units of uniquely bound features that compose the conscious representation.

**Simulated**
Patterns of neural activity can be generated intrinsically from the internal dynamics of the brain. Higher-order current generators act as simulators, predicting the incoming neural signal through the production of a phase shift in ongoing electrical patterns sent back early processing areas to act as a local pacemaker of neural activity (Bollimunta, Chen, Schroeder, & Ding; 2008; Bullier, 2001). This aspect of brain functioning has been modeled computationally as pattern completion inference, where the top-down prediction generates missing stimuli from partial input to the network (O'Reilly, 2017).

**Situated**
These recurrent electrophysiological feedback mechanisms between brain regions are accomplished through a system of nested oscillations in distinct frequency bands of activity spanning many orders of magnitude (Buzsaki, 2006). S3Q contends that qualia are situated in the conscious representation through this synchronized, hierarchical embedding code that clusters the computational units relative to each other (Orban, 2008). Mechanisms of this clustering, such as lateral inhibition where an activated neuron will inhibit the neighboring neurons for preferential selection, can be explored in modeling experiments implementing features such as a k-winner-take-all algorithm (Kriete, Noelle, Cohen, and O'Reilly, 2013). One reasonable theoretical model suggests reentrant synchronization enables the unique properties of the conscious representation in humans, such as the binding of color, shape, and location features (Seth, McKinstry, Edelman, and Krichmar, 2004). Computational models implementing pattern completion accomplish binding of visual stimuli by predicting the object and features to be recognized (Majani, Erlanson, and Abu-Mostafa, 1988; Kriete et al., 2013; Bressler, Tang, Sylvester, Shulman, and Corbetta, 2008). Combining the ideas of clustering and prediction for binding might be represented operationally by a classic Hopfield network/Boltzmann machine constraint satisfaction system where top-down predictions lead to contrast enhancement through mutual excitation which results in a winning set of neurons encoding the relevant features and the inhibition of irrelevant features (Van Rullen, 2000).

**Structurally Coherent**
Groups of neurons that are engaged in synchronous neural oscillations have temporal communication windows of excitability based on the phase of the activity waveform. During the 'Up' phase of the oscillation, the network is able to communicate, while during the 'Down' state the network is decoupled from sensory input. Only oscillations in phase synchrony can effectively interact, because the communication windows are open at the same time. The temporal difference created from this predicted/actual outcome can provide an error measure that can be used for predictive error driven learning in a simple recurrent network (O'Reilly, 2017). It is hypothesized that structural coherence can be indexed as a function of this error measure. Phase synchrony across distinct oscillatory bands provides a structure of brain-wide communication whereby faster rhythms reflect local communication and slower rhythms afford long distance communication across the network (Buszaki, 2006). This pattern of structural coherence across the distinct frequency bands combined with an internally generated simulated and situated world-model dictates the flexible structure of human cognition (Fries, 2005).

**CONCLUSIONS**
Flexibility means that an agent can operate successfully when conditions are variable, unexpected, or when novel tasks arise. S3Q suggests that human conscious representations accomplish this desired flexibility through the relational structure of synchronizing neural networks in hierarchical representations. These conscious representations overcome practical limitations such as partial observability through a suitable balance of simulation, situatedness, and structural coherence. These three characteristics are distinct, but closely interrelated. S3Q theory conjectures that agents implementing a representation suitably balancing S3Q tenets are afforded flexible responding to novel queries. The phenomenal 'what it is like' characteristics of this conscious representation will be recognized in the specific configuration of the simulated, situated, and structurally coherent qualia.